\documentclass[twocolumn,showpacs,preprintnumbers,amsmath,amssymb]{revtex4}
\usepackage{graphicx}
\usepackage{dcolumn}
\DeclareMathAlphabet{\mathpzc}{OT1}{pzc}{m}{it}

\newcommand{\ltsim}{\mbox{{\raisebox{-0.4ex}{$\stackrel{<}{{\scriptstyle\sim
}}
$}}}}
\begin{document}

\title{A de Haas-van Alphen study of the 
filled skutterudite
compounds PrOs$_4$As$_{12}$ and LaOs$_4$As$_{12}$}

\author{Pei-Chun Ho$^1$, J. Singleton$^2$, M.B. Maple$^1$,
Hisatomo Harima$^3$, P.A. Goddard$^{2,4}$, Z. Henkie$^5$
and A. Pietraszko$^5$}

\affiliation{$^1$Department of Physics and
Institute for Pure and Applied Physical Sciences,
University of California, San Diego, 9500 Gilman Drive,
Dept. 0360, La Jolla, CA~92093-0360.}
\affiliation{$^2$National High Magnetic Field Laboratory, 
Los Alamos National Laboratory, MS-E536, 
Los Alamos, New Mexico 87545, USA.}
\affiliation{$^3$Department of Physics, 
Kobe University, 1-1 Rokko-dai Noda Kobe 657-8501, JAPAN}
\affiliation{$^4$Clarendon Laboratory, 
Oxford University, Parks Road, Oxford, OX1 3PU, UK}
\affiliation{$^5$Institute for Low Temperature and 
Structure Research, Polish Academy of Sciences,
50-950, Wroclaw, POLAND}

%\date{today}

\begin{abstract}
Comprehensive magnetic-field-orientation dependent 
studies of the susceptibility and
de Haas-van Alphen effect have
been carried out on single crystals of the 
filled skutterudites PrOs$_4$As$_{12}$ and
LaOs$_4$As$_{12}$ using magnetic fields
of up to 40~T. Several peaks are observed
in the low-field susceptibility of PrOs$_4$As$_{12}$,
corresponding
to cascades of metamagnetic transitions
separating the low-field antiferromagnetic
and high-field paramagnetic metal (PMM) phases.
The de Haas-van Alphen experiments show
that the Fermi-surface topologies of
PrOs$_4$As$_{12}$ in its PMM phase
and LaOs$_4$As$_{12}$ are
very similar. In addition, they are
in reasonable agreement with the predictions 
of bandstructure calculations for LaOs$_4$As$_{12}$
on the PrOs$_4$As$_{12}$ lattice.
Both observations suggest that
the Pr 4$f$ electrons contribute little to the
number of itinerant quasiparticles in the PMM phase. 
However, whilst the properties of LaOs$_4$As$_{12}$
suggest a conventional nonmagnetic Fermi liquid,
the effects of direct exchange and electron
correlations are detected in the PMM phase of PrOs$_4$As$_{12}$.
For example, the quasiparticle effective masses
in PrOs$_4$As$_{12}$ are found to decrease with 
increasing field, probably reflecting the gradual suppression of
magnetic fluctuations associated with proximity to the low-temperature,
low-field antiferromagnetic state.
\end{abstract}

\maketitle
\section{Introduction}
Filled skutterudite compounds, with the formula
MT$_4$X$_{12}$, where  M is
an alkali metal, alkaline-earth, lanthanide, or actinide,
T is Fe, Ru, or Os and X is P, As, or Sb, display a wide
variety of interesting phenomena caused by
strong electron 
correlations~\cite{aokireview,oneprime,meisner}.
The case M = Pr has attracted particular interest,
where it is thought that 
many of the physical properties are
attributable to the
ground state and the low-lying excited state of the Pr$^{3+}$ ion in
the crystalline electric field  (CEF), and the hybridization of
the Pr 4f orbitals with the ligand states of the surrounding
ionic 
cage~\cite{kuric,aokimusr,twoprime,sugawaraprfep,sugawaraprossb,fourprime}.
A variety of correlated electron phenomena have
been observed in the Pr-based filled skutterudites: conventional
BCS  and unconventional superconductivity, magnetic
order, quadrupolar order, metal-insulator transitions,
Kondo phenomena, heavy-fermion behavior, and non-Fermi-liquid
behavior~\cite{aokireview,oneprime,kuric,aokimusr,twoprime,sugawaraprfep,fourprime,maplesupercond,sugawaratspt}.
Much of this work has been carried out on the
Pr-based filled skutterudite phosphides and antimonides;
by contrast, the arsenides have not been investigated in much
detail~\cite{yuhaszprb,maplepnas}.
In this paper, we
present an investigation of the de Haas-van Alphen
effect and susceptibility in PrOs$_4$As$_{12}$ single crystals.
Analagous experiments were also carried out
on crystals of
the isostructural nonmagnetic metal LaOs$_4$As$_{12}$.

Previously, PrOs$_4$As$_{12}$ has been shown to enter an
antiferromagnetic state at low temperatures;
this phase possesses a greatly enhanced
electronic specific heat coefficient $\gamma \approx 1$~Jmol$^{-1}$K$^{-2}$,
highly suggestive of heavy-fermion behavior~\cite{yuhaszprb}.
On applying magnetic fields $\mu_0 H \approx 3-4$~T, 
the antiferromagnetic
phase is destroyed and the value of $\gamma$ collapses~\cite{maplepnas}.
The measurements presented in the current paper 
show that this antiferromagnetic to paramagnetic metal
transition in fact proceeds via 3 or 4 metamagnetic transitions,
the field positions of which depend strongly on the orientation
of the crystal in the field.

The de Haas-van Alphen experiments show that
the Fermi surface topologies of
PrOs$_4$As$_{12}$ in its paramagnetic metal phase
and LaOs$_4$As$_{12}$
are rather similar.
However, whereas LaOs$_4$As$_{12}$ behaves as a fairly conventional
low-effective-mass Fermi liquid, we find that 
the paramagnetic metal phase of PrOs$_{4}$As$_{12}$
exhibits the effects of a direct exchange coupling
and shows quasiparticle effective masses that are renormalized
by fluctuations associated with the proximity of the
low-field, low-temperature antiferromagnetic state. 
\section{Experimental details and bandstructure calculations}
Single crystals of PrOs$_4$As$_{12}$ and LaOs$_4$As$_{12}$
were grown from elements with purities 99.9\%
using the high-temperature molten-metal-flux
procedure described in Ref.~\cite{yuhaszprb}.
As-grown crystals were cleaned
in acid to remove residual flux and
impurity phases from their
surfaces.
The crystals (cubic space group Im$\bar{3}$;
for structural details see Refs.~\cite{maplepnas,yuhaszprb})
are truncated octahedra 
(8 large $\{111\}$ faces and 6, approximately
square, smaller
$\{100\}$ faces), with largest dimensions in the
range $0.1-1$~mm; the presence of
well-defined crystal faces greatly aids the
accurate orientation of the samples in the
magnetic field.

The measurements carried out in quasistatic magnetic fields employ
a torque magnetometer with a cantilever constructed from $5~\mu$m
phosphor bronze~\cite{brooks}. A single crystal is glued
to the cantilever via a thin sheet
of strain-reducing paper. The interaction
between the sample's magnetic moment {\bf m}
and the applied magnetic field {\bf B}
causes a torque $-{\bf m} \times {\bf B}$
that results in a deflection of the cantilever.
The deflection is monitored using the capacitance
between the cantilever and a fixed plate about 1~mm below it
($\sim 0.5$~pF), measured using a General Radio Capacitance
bridge.
Care is taken to ensure that deflections are small,
so that the sample's orientation in the field is
not changed significantly by the torque.
The torque magnetometer is mounted on a two-axis
cryogenic goniometer that allows the sample orientation
to be changed {\it in situ}; $^3$He refrigeration
provides temperatures in the range $0.45-10$~K.
Quasistatic magnetic fields were provided by a
superconducting magnet in Los Alamos and by 
33~T Bitter coils at NHMFL Tallahassee.

Pulsed-field experiments use
a 1~mm bore, 1~mm long compensated-coil susceptometer,
constructed from 50-gauge high-purity copper wire.
The coil is wound with approximately 640 turns
in one sense, followed by around 360 in the opposite
sense; final turns are added or subtracted
by hand on the bench-top to reduce the uncompensated
area of the coil to a fraction of a turn.
Fine-tuning of the compensation is accomplished
by electronically adding or subtracting a small part of the
voltage induced in a coaxial single-turn coil wound
around the susceptometer~\cite{neilandme}.
Once this has been done,
the signal from the susceptometer is
$V\propto ({\rm d}M/{\rm d}t) = ({\rm d}M/{\rm d}H)({\rm d}H/{\rm d}t)$,
where $M$ is the magnetization of a sample
placed within the bore of the coil
and $H$ is the applied magnetic field~\cite{neilandme}.
Magnetic fields were provided by the
40~T mid-pulse magnet at NHMFL Los Alamos~\cite{nhmfl};
the use of this magnet, with its relatively slow
rise time ($\approx 100$~ms) was necessary to
avoid inductive sample heating.
The susceptometer was placed within a
simple $^3$He cryostat providing temperatures
down to 0.4~K.
Magnetic fields were deduced by integrating the
voltage induced in an eleven-turn coil,
calibrated by observing the de Haas-van Alphen
oscillations of the belly orbits of
the copper coils of the susceptometer~\cite{neilandme}.

The bandstructure of PrOs$_{4}$As$_{12}$ was calculated
using a FLAPW and LDA method
(for further details see Refs.~\cite{sugawaraprossb,ha19,ha20}).
The 4$f$ electrons are assumed to be localized,
so that the calculation is essentially for
LaOs$_4$As$_{12}$ on the PrOs$_4$As$_{12}$ lattice.
\section{Low-field susceptibility and phase diagram}
\begin{figure}[t]
\includegraphics[width=7.5cm]{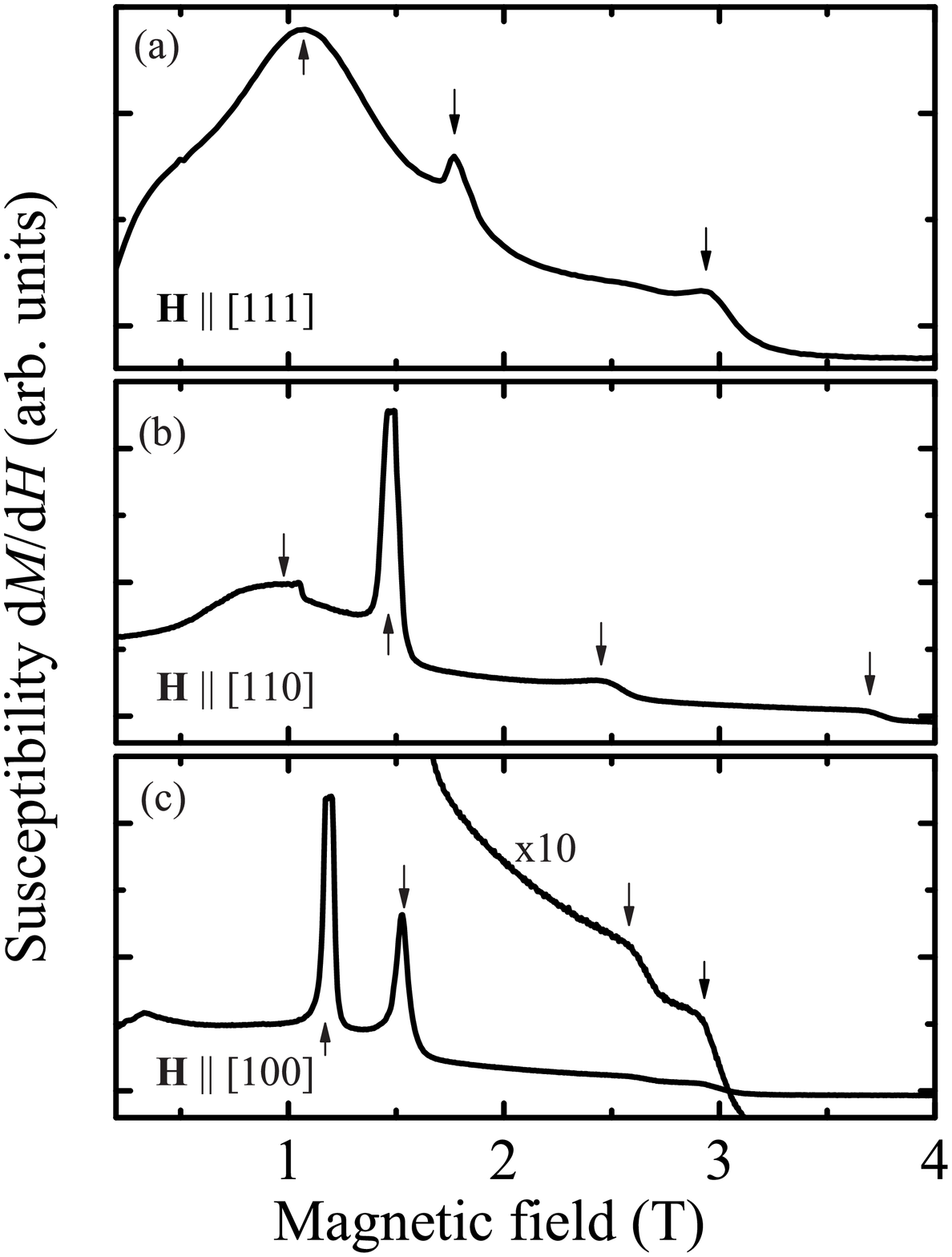}
\caption{(a)~Differential susceptibility
$({\rm d}M/{\rm d}H)$ 
of a single crystal of PrOs$_4$As$_{12}$
for ${\bf H} || [111]$ ($T=0.57$~K).
Three metamagnetic transitions are observed as
peaks, indicated by arrows. It is possible
that a fourth
transition occurs at about 2.6~T.
(b)~Similar susceptibility data for
${\bf H}||[110]$ ($T=0.51$~K);
four metamagnetic transitions may be seen
(arrows).
(c)~Susceptibility data for
${\bf H}||[100]$ ($T=0.51$~K);
the data between 2 and 3~T have been
multiplied by 10 to enhance the visibility of features.
Four metamagnetic transitions may be seen
(arrows).}
\label{fig1}
\vspace{-5mm}
\end{figure}

\begin{figure}[t]
\includegraphics[width=9.5cm]{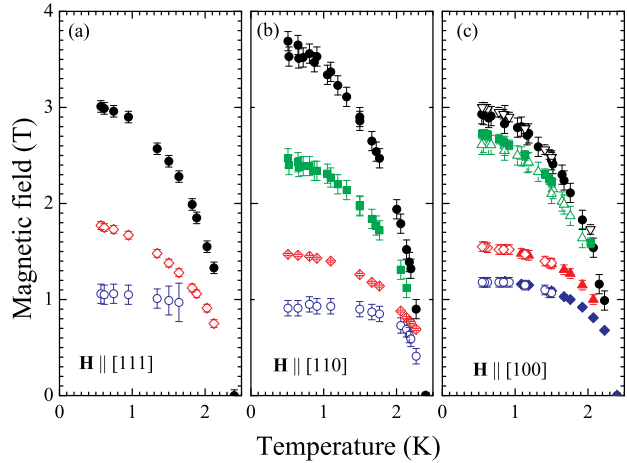}
\caption{$H,T$ phase diagrams of the peaks in
the susceptibility of PrOs$_4$As$_{12}$
for (a)~${\bf H}||[111]$
(b)~${\bf H}||[110]$ and (c)~${\bf H}||[100]$.
Data for samples from two different batches are shown
in (c) (indicated by filled and hollow symbols), 
indicating that the peak positions are an intrinsic
feature of the material.
The transitions surround the
low-field, low-temperature
antiferromagnetic phase~\cite{maplepnas};
the region outside the transitions is
the paramagnetic metal phase.}
\label{fig2}
\end{figure}

Fig.~\ref{fig1} shows the low-field
susceptibility $({\rm d}M/{\rm d}H)$
of a single crystal of PrOs$_4$As$_{12}$,
measured using the pulsed-field susceptometer,
for three different orientations
of the sample in the magnetic field.
Previous heat-capacity experiments for 
${\bf H}||[111]$~\cite{yuhaszprb,maplepnas}
have suggested that two field-induced phase transitions
occur in PrOs$_4$As$_{12}$ at low temperatures. However,
in the present experiments, at each orientation,
three or four peaks (or humps) are observed in $({\rm d}M/{\rm d}H)$,
corresponding to metamagnetic transitions involving
a broadened, step-like change 
in $M$ (see e.g., Ref.~\cite{goddardsteps}). 
The $(H,T)$ positions of the transitions are plotted in
Fig.~\ref{fig2}; the $H=0$ point is
$T\approx 2.4$~K, a temperature above which there
is definitely no discernable trace of features in the 
low-field susceptibility. This is
close to the zero-field N\'{e}el temperatures
$T_{\rm N}=2.28$~K and $T_{\rm N}=2.5$~K
inferred respectively from neutron-scattering
and conventional magnetometry experiments 
on PrOs$_4$As$_{12}$~\cite{maplepnas}.
This suggests that the low-field antiferromagnetic phase is a
prerequisite for the transitions seen in the susceptibility.
The fact that the transitions are closely spaced and
occur at fields that change rather slowly with changing temperature
may be the reason why fewer features have been detected in
fixed-field, swept temperature experiments~\cite{maplepnas}.

By contrast, apart from a feature attributable
to the superconducting critical field~\cite{shirotani},
the low-field susceptibility of
LaOs$_4$As$_{12}$ was free of metamagnetic transitions,
as might be expected for what is thought to be a relatively
conventional, nonmagnetic Fermi liquid~\cite{shirotani}.

\section{de Haas-van Alphen effect}
\subsection{Frequency spectrum and the effects of direct exchange}
\begin{figure}[t]
\includegraphics[width=9.5cm]{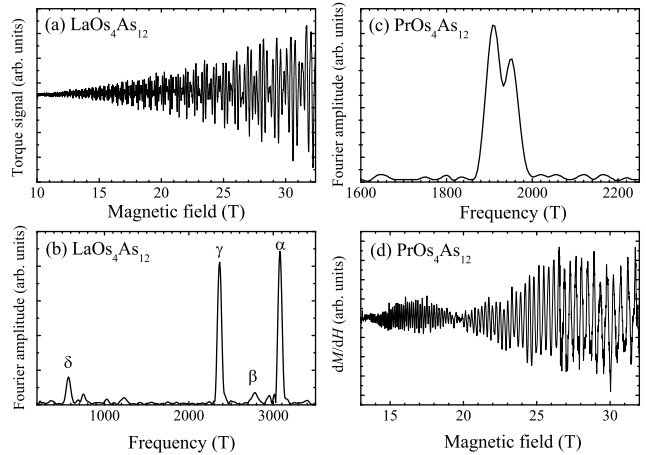}
\vspace{-4mm}
\caption{(a)~de Haas-van Alphen oscillations in
the torque magnetometer signal for LaOs$_4$As$_{12}$ 
($T=1.0$~K). The quantity plotted is the change
in capacitance of the magnetometer, proportional
to the torque for small 
deflections~\cite{brooks}. The sample
has been tilted about the $[1\bar{1}0]$ axis
so that its $[100]$ direction makes an angle
of $\theta = 15.3^{\circ}$ with the magnetic field.
(b)~Fourier transform of the data in (a),
showing four de Haas-van Alphen frequencies,
labelled $\alpha-\delta$.
(c)~Expansion of a Fourier transform
of PrOs$_4$As$_{12}$ de Haas-van Alphen
oscillations (pulsed-field susceptometer)
close to the $\gamma$ frequency.
The field is parallel to $[110]$, the
temperature is 0.52~K and a 
field window of $12-30$~T was used for
the transform in order to attain a high
resolution. Under such conditions
it can be seen that the $\gamma$
peak is split; i.e., the $\gamma$
oscillations in PrOs$_4$As$_{12}$
in fact comprise two closely-spaced
frequencies $\gamma_1$ and $\gamma_2$. 
A similar splitting is seen for the
$\alpha$, $\beta$ and $\delta$ frequencies.
(d)~Pulsed-field susceptometer data
for PrOs$_4$As$_{12}$ (0.52~K) showing
a node at 20~T caused by the beating of
the $\gamma_1$ and $\gamma_2$ frequencies.}
\label{dhva}
\end{figure}

At fields above those shown in Figs.~\ref{fig1} and
\ref{fig2} (i.e., in the paramagnetic metal
phase of PrOs$_4$As$_{12}$), the low-temperature responses 
of the torque magnetometer
and pulsed-field susceptometer become dominated by
de Haas-van Alphen oscillations.
An example of oscillations in
the torque is shown in Fig.~\ref{dhva}(a) for
LaOs$_4$As$_{12}$.
The complex form of the oscillations
suggests the presence of 
several de Haas-van Alphen frequencies,
each corresponding, via the Onsager relationship, 
to different Fermi-surface
cross-sections in the plane perpendicular to the 
magnetic field~\cite{shoenberg}.
Numerical Fourier transformation was used to
extract the frequencies of the oscillations (Fig.~\ref{dhva}(b)). 
Once obvious higher harmonics are accounted
for in the Fourier transforms, 
four de Haas-van Alphen frequencies are
consistently observable in PrOs$_4$As$_{12}$
and LaOs$_4$As$_{12}$ (Fig.~\ref{dhva}(b)),
corresponding to four Fermi-surface sections.  
Following the precedents set by de Haas-van Alphen
studies of other filled 
skutterudites~\cite{sugawaraprfep,sugawaraprossb,sugawaraold},
we label the Fermi-surface sections (frequencies) 
using the Greek letters $\alpha-\delta$, with $\alpha$
corresponding to the largest orbit (Fig.~\ref{dhva}(b)).

Although the frequencies observed in PrOs$_4$As$_{12}$
and LaOs$_4$As$_{12}$ are similar, an interesting difference
between the materials is visible if a large enough field
window (i.e., high enough resolution~\cite{shoenberg})
is used for the Fourier transform of the data.
Whereas in LaOs$_4$As$_{12}$ each of the peaks
$\alpha-\delta$ in Fourier amplitude
(Fig.~\ref{dhva}(b)) corresponds to
a single frequency, those in PrOs$_4$As$_{12}$
each comprise a pair of closely-spaced frequencies
(Fig~\ref{dhva}(c)). These very similar frequencies
cause a beating in the de Haas-van Alphen
signal, resulting in observable nodes;
an example is shown in Fig.~\ref{dhva}(d).
Some illustrative values of the splitting are
given in Table~\ref{table1}.

The presence of pairs of closely-spaced frequencies
suggests that each of the Fermi-surface sections
in PrOs$_4$As$_{12}$ is split into spin-up and spin-down
components by direct exchange coupling, as befits
a polarized, strongly paramagnetic metal~\cite{goodrich}.
The exchange interactions act
as an effective field that results in
populations of spin-up and spin-down quasiparticles
that are no longer equal. Thus, the Fermi-surface
cross-sections 
differ for spin-up and spin-down quasiparticles,
resulting in the observed splitting of each de Haas-van
Alphen frequency into two closely-spaced components.
This effect has been noted in a number of
magnetic hexaborides (see Ref.~\cite{goodrich} and citations
therein).
Later in the paper we shall use the frequency splitting and
other data to estimate the exchange coupling in
the paramagnetic phase of PrOs$_4$As$_{12}$.

\subsection{Quasiparticle effective masses}
Effective masses were are derived 
from the de Haas-van Alphen oscillations by
fitting the temperature-dependent Fourier
amplitude $A(T)$ to the
relevant part of the three-dimensional
Lifshitz-Kosevich formula~\cite{shoenberg}
\begin{equation}
A(T) \propto \frac{\chi}{\sinh \chi},
\label{LKmass}
\end{equation}
where $\chi = 14.69 m^*T/B$,
with $m^*$ the quasiparticle cyclotron effective mass
and $B$ the reciprocal of the mean inverse
field of the window used for the Fourier transform.

First, the effective masses corresponding 
to spin-up and spin-down components of
each de Haas-van Alphen frequency in PrOs$_4$As$_{12}$
were evaluated independently, using a wide Fourier
window to ensure that the two separate
components were resolved (see Fig~\ref{dhva}(c)).
Within experimental errors, the effective masses
of the spin-up and spin-down components of
each frequency were found to be identical.

The fact that both spin-up and spin-down
components have very similar masses is an
important consideration in the
investigation of the field dependence
of the effective masses in PrOs$_4$As$_{12}$
described in the following paragraphs.
Such a study necessitates the use of a narrower
field window (and thus lower frequency
resolution); hence, it
is not possible to resolve the separate
spin-up and spin-down components
of each Fermi-surface section.
The masses evaluated are therefore
the average effective masses of the
spin-up and spin-down components of each 
Fermi-surface section.

\begin{figure}[t]
\includegraphics[width=8cm]{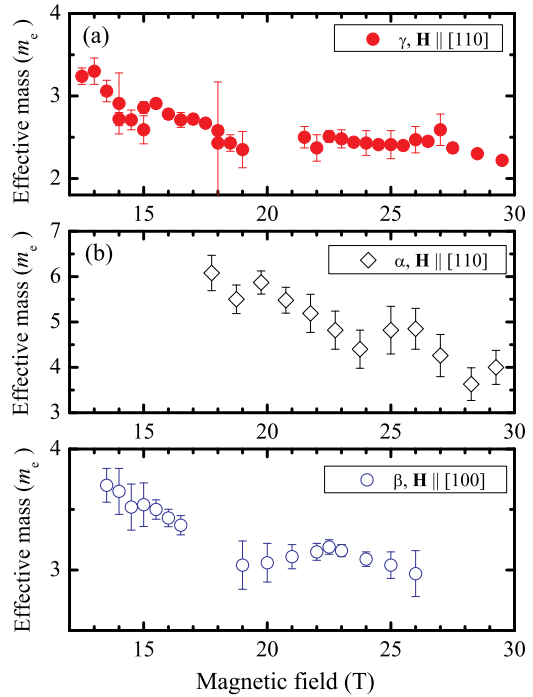}
\caption{Examples of field-dependent
effective masses of PrOs$_4$As$_{12}$ 
(paramagnetic metal phase)
derived from the
temperature dependence of the de Haas-van Alphen
oscillations (Eq.~\ref{LKmass}). 
The magnetic field plotted is the
reciprocal of the mean inverse field of
the Fourier-transform window; typically,
the window width used was in the range $0.005-0.011$~T$^{-1}$.
(a)~$\gamma$ Fermi-surface section, ${\bf H}||[110]$.
(b)~$\alpha$ Fermi-surface section, ${\bf H}||[110]$.
(c)~$\beta$ Fermi-surface section, ${\bf H}||[100]$.
Note that gaps in the data correspond to
nodes in the de Haas-van Alphen oscillations
caused by the beating of the spin-up and spin-down
components (see Fig.~\ref{dhva}(c),(d)).}
\label{fig4}
\end{figure}

To investigate the field dependence of
the effective masses, 
de Haas-van Alphen oscillations were recorded
at several fixed temperatures in the range
$0.45-3.0$~K.
For each data set, Fourier transforms were performed over
a restricted window of inverse magnetic field (typical window
width $0.005-0.011~{\rm T}^{-1}$),
and the temperature-dependent amplitude fitted
to Eq.~\ref{LKmass}.
Typical mass data are shown in Fig.~\ref{fig4}.
Within the errors of the experiment,
the masses increase with decreasing magnetic field,
with the decrease becoming more noticeable below $\mu_0H=20$~T.

The possibility of magnetic-field dependent effective
masses in strongly-correlated electron systems has
been discussed by a number of authors~\cite{wassspring,edwardsgreen,schotte}.
For instance, Ref.\cite{schotte} uses a Kondo-impurity model
that invokes a narrow resonance 
(width $\sim k_{\rm B}T_{\rm K}$, 
where $T_{\rm K}$ is the Kondo temperature) 
in the density of states at the chemical potential that undergoes 
Zeeman splitting in a magnetic field.
Though this approach is only strictly
applicable to dilute alloys such as La$_{1-x}$Ce$_x$B$_6$ 
($x \ll 1$~\cite{mihut}),
it should be noted that it can provide a 
phenomenological description of the heat capacity
$C(H,T)$ of the paramagnetic phase of
PrOs$_4$As$_{12}$ for $\mu_0H~ \ltsim~16$~T
and $0.4~\ltsim~ T~\ltsim 12$~K~\cite{maplepnas}.  
The model predicts a strong decrease 
in the electronic specific heat coefficient $\gamma$ (and, consequently, 
the average quasiparticle effective mass) with magnetic field,
yielding a small Kondo 
temperature $T_{\rm K} \sim 3.5$~K. 
This is similar to the value $T_{\rm K}\sim 1$~K 
obtained from a scaling analysis of the electrical 
resistivity $\rho(H,T)$ as a function of $H$ and $T$~\cite{maplepnas}.
The scaling analysis was 
based on a calculation due to Schlottmann~\cite{schlott}, 
also for the Kondo impurity case. 
The two models applied to 
the $C(H,T)$ and $\rho(H, T)$ data for PrOs$_4$As$_{12}$ 
in the paramagnetic phase, were for the case of spin $S = 1$, 
which is appropriate for a triplet ground state ($\Gamma_5$) for 
PrOs$_4$As$_{12}$ in the crystalline 
electric field~\cite{maplepnas,yuhaszprb}.

In spite of this qualitative success
of the Kondo impurity models,
the Anderson lattice models used
to treat intinerant heavy-fermion 
metals~\cite{wassspring,edwardsgreen}
might be expected to be more relevant to PrOs$_4$As$_{12}$.
However, Refs.~\cite{wassspring,edwardsgreen} 
suggest very different behavior for
the spin-up and spin-down components of each Fermi-surface section,
a prediction that was backed up at least qualitatively by
the experiments on pure
CeB$_6$~\cite{springford} and CePd$_2$Si$_2$~\cite{sheikin}. 
In CePd$_2$Si$_2$, differing, field-dependent masses are inferred
for the two spin components of the Fermi surface~\cite{sheikin}.
However, in CeB$_6$, a field-dependent mass
is observed for the minority-spin Fermi-surface
section whilst the de Haas-van Alphen
oscillations for the majority-spin Fermi surface are not observed,
because of a very heavy mass 
and/or spin-dependent scattering~\cite{goodrichb6,teklu}.
(It should be noted that the situation
in Ce$_x$La$_{1-x}$B$_6$ with large Ce
concentrations is less clear cut, with some
studies reporting only one spin component
in the de Haas-van Alphen oscillations~\cite{goodrichb6,teklu},
whilst others infer the presence of spin-up and spin-down
Fermi-surface sections from detailed analysis of the
temperature dependence or phase of the 
oscillations~\cite{mihut,endo}.)

By contrast with the predictions of
Refs.~\cite{wassspring,edwardsgreen}, 
both spin-up and spin-down components
are observed for all Fermi-surface sections of PrOs$_4$As$_{12}$,
and their behavior with changing field
(e.g., the values of their effective masses) appears identical.
It is therefore unlikely that the 
Anderson lattice models~\cite{wassspring,edwardsgreen}
are applicable to PrOs$_4$As$_{12}$. 
Instead, the enhancement of the effective mass
in the paramagnetic metal phase of PrOs$_4$As$_{12}$
is most likely to be due to fluctuations associated
with the proximity of the antiferromagnetic phase.
As one moves away from this phase in magnetic field,
the fluctuations will be gradually suppressed,
leading to quasiparticle masses that gradually
decrease (Fig.~\ref{fig4}).
The gradual reduction in
in the electronic contribution to the
heat capacity with increasing field
in the paramagnetic metal phase referred to 
above~\cite{maplepnas};
is likely to be due to the same mechanism.

Some illustrative values of the effective masses
in PrOs$_4$As$_{12}$
for the various Fermi-surface sections, evaluated at
a field of 25~T, are shown in Table~\ref{table1}.
By contrast, the behavior of LaOs$_4$As$_{12}$
is more conventional, in that the masses
are relatively small compared to those in PrOs$_4$As$_{12}$, 
and field-independent; a summary
is given in Table~\ref{table2} (c.f. the comparison of
PrOs$_4$Sb$_{12}$ and LaOs$_4$Sb$_{12}$ given in
Ref.~\cite{sugawaraprossb}).
Effective masses associated with various Fermi-surface orbits
may also be deduced from the LDA/FLAPW calculations
(the identification of the orbits involved is
covered in the following section);
theoretical values are compared with experimental
data in Table~\ref{table3}.
The fluctuations discussed above
lead to experimental masses at 25~T that are enhanced
by a factor $\sim 3-5$ with respect to the
bandstructure calculations.

Finally, the effective masses can be used to
estimate the direct exchange energy in PrOs$_4$As$_{12}$.
This is done by comparing effective Fermi energies
for the spin-up and spin-down components.
For a particular de Haas-van Alphen
frequency $F$, with associated effective mass $m^*$,
the effective Fermi energy is $\hbar e F/m^*$.
The direct exchange energy $E_{\rm exch}$
is one half
of the difference between effective Fermi energies
for the spin-up and spin-down components~\cite{goodrich}:
\begin{equation}
E_{\rm exch}=\frac{\hbar\Delta F}{2m^*},
\label{hopeless}
\end{equation}
where $\Delta F$
is the difference in de Haas-van Alphen frequency.
Some illustrative values of $E_{\rm exch}$
are listed in Table~\ref{table1};
all are close to 1~meV, and similar 
in magnitude to the exchange
energies observed in the 4f hexaborides (see Ref.~\cite{goodrich}
and citations therein).

\begin{table}[tbp] 
\centering
\begin{tabular}{|l|l|l|l|l|l|}
\hline
 FSS  & Orient'n & $<F>$ (T) & $\Delta F$ (T) & $m^*~(m_{\rm e})$ & $E_{\rm exch}$~(meV) \\ 
\hline
$\gamma$ & ${\bf H}||[100]$ & $2210\pm 10$ & 50 & $2.6 \pm 0.2$ & 1.1 \\
$\beta$  & ${\bf H}||[100]$ & $2740\pm 10$ & 75 & $3.0 \pm 0.1$ & 1.4 \\
$\alpha$ & ${\bf H}||[100]$ & $3005\pm 10$ & 60 & $4.4 \pm 0.1$ & 0.8 \\
\hline
$\gamma$ & ${\bf H}||[110]$ & $1926\pm 10$ & 40 & $2.4 \pm 0.2$ & 1.0 \\
$\alpha$ & ${\bf H}||[110]$ & $3510\pm 10$ & 60 & $4.8 \pm 0.3$ & 0.7 \\
\hline
$\gamma$ & ${\bf H}||[111]$ & $2028\pm 10$ & 50 & $2.7 \pm 0.4$ & 1.1 \\
\hline
\end{tabular}
\caption{Selected Fermi-surface properties of PrOs$_4$As$_{12}$
in its paramagnetic metal phase.
The first column denotes the Fermi-surface section (FSS),
and the second the field orientation; $<F>$ is the mean
of the separate spin-up and spin-down frequencies,
and $\Delta F$ is their difference, evaluated at $\mu_0H = 25$~T. 
The effective masses are also
evaluated at $25$~T, using a Fourier window of width $0.011$~T$^{-1}$.
The final column is the estimated direct exchange energy
calculated using the tabulated parameters.}
\label{table1}
\end{table}

\begin{table}[tbp] 
\centering
\begin{tabular}{|l|l|l|l|}
\hline
 FSS  & Orient'n & $F$ (T) & $m^*~(m_{\rm e})$ \\ 
\hline
$\gamma$ & ${\bf H}||[100]$ & $2425\pm 10$ & $1.31 \pm 0.08$ \\
$\alpha$ & ${\bf H}||[100]$ & $3045\pm 10$ & $2.0 \pm 0.1$ \\
\hline
$\gamma$ & ${\bf H}||[110]$ & $2182\pm 10$ & $1.12 \pm 0.04$ \\
$\alpha$ & ${\bf H}||[110]$ & $3770\pm 10$ & $3.3 \pm 0.5$ \\
\hline
$\gamma$ & ${\bf H}||[111]$ & $2185\pm 10$ & $1.14 \pm 0.05$ \\
$\alpha$ & ${\bf H}||[111]$ & $3525\pm 10$ & $2.8 \pm 0.2$ \\
\hline
\end{tabular}
\caption{Selected Fermi-surface properties of LaOs$_4$As$_{12}$.
The first column denotes the Fermi-surface section (FSS),
and the second the field orientation; $F$ is the 
de Haas-van Alphen frequency.
The effective masses, which are field-independent, are
evaluated at $13.6$~T.}
\label{table2}
\end{table}

\begin{table}[tbp] 
\centering
\begin{tabular}{|l|l|l|l|l|l|}
\hline
 FSS  & $F_{\rm expt}$ (T) & $F_{\rm theory}$ (T) & 
$|\frac{m^*_{\rm expt}}{m_{\rm e}}|$
& $\frac{m^*_{\rm theory}}{m_{\rm e}}$ & 
$|\frac{m^*_{\rm expt}}{m^*_{\rm theory}}|$ \\ 
\hline
${\bf H} || [100]$ & & & & & \\ 
$\gamma$ & 2210 & 2245 & $2.6\pm 0.2$ & -0.86 & 3.0 \\
$\beta$  & 2740 & 2621 & $3.0 \pm 0.1$ & -0.87 & 3.4 \\
$\alpha$ & 3005 & 2651 & $4.4 \pm 0.1$ & -0.95 & 4.6 \\
\hline
\hline
${\bf H} || [110]$ & & & & & \\ 
$\delta$ & 550 & 775 & -  & -1.47 & - \\
$\gamma$  & 1926 & 1932 & $2.4 \pm 0.2$ & -0.61 & 3.9 \\
$\alpha$ & 3510 & 3329 & $4.8 \pm 0.3$ & -1.60 & 3.0 \\
\hline
\hline
${\bf H} || [111]$ & & & & & \\ 
$\gamma$  & 1932 & 2028 & $2.7 \pm 0.4$ & -0.60 & 4.5 \\
\hline
\end{tabular}
\caption{Experimental (subscript ``expt'';
paramagnetic metal phase)
and theoretical (subscript ``theory'')
de Haas-van Alphen frequencies $F$
and effective masses $m^*$
for PrOs$_4$As$_{12}$. The experimental
masses were evaluated
at 25~T; see Table~\ref{table1}.}
\label{table3}
\end{table}

\subsection{Three-dimensional Fermi-surface topology}
\begin{figure}[t]
\includegraphics[width=8.5cm]{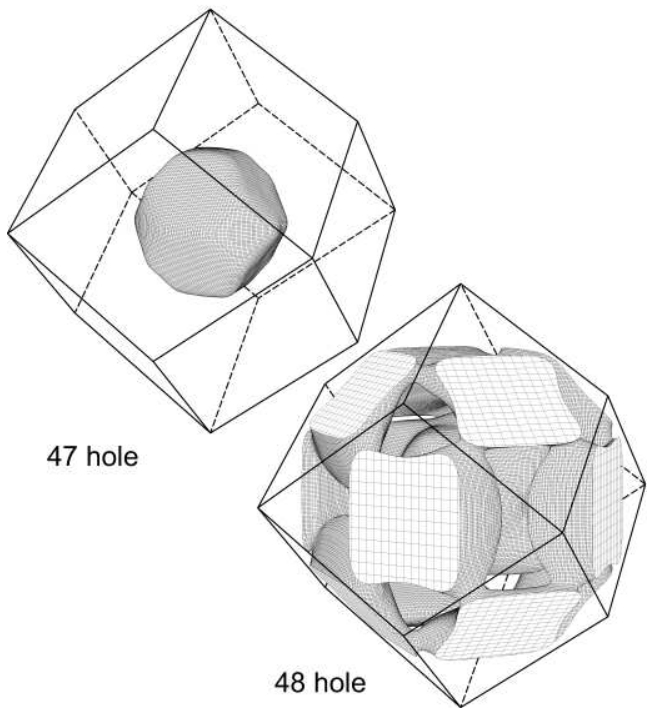}
\caption{Predicted Fermi surface of
PrOs$_4$As$_{12}$, using the LDA/FLAPW
calculation of LaOs$_4$As$_{12}$ on the PrOs$_4$As$_{12}$
lattice. For clarity, the $47^{\rm th}$ and $48^{\rm th}$
hole-band Fermi-surface sections are shown separately.
}
\label{fermifig}
\end{figure}

The Fermi surface of PrOs$_4$As$_{12}$
predicted by the LDA/FLAPW calculations is shown 
in Fig.~\ref{fermifig}. 
In order to check this three-dimensional
picture of the Fermi-surface,
torque magnetometry data sets were recorded
at $^3$He base temperature
for both PrOs$_4$As$_{12}$ and LaOs$_4$As$_{12}$
at several orientations of the magnetic field,
using the cryogenic goniometer to rotate the
sample. A disadvantage of
torque magnetometry is that the signal
is very small or absent when the field is exactly
aligned along a symmetry direction of the crystal;
hence, additional data were recorded
with the field along the $[111]$, $[110]$
and $[100]$ directions using the pulsed-field
susceptometer. Both techniques are in
good agreement.

\begin{figure}[t]
\includegraphics[width=9.5cm]{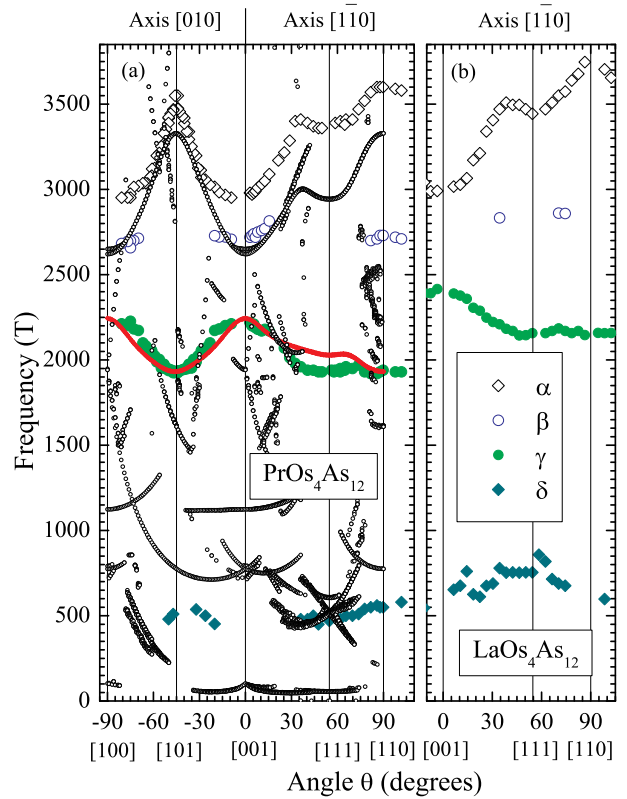}
\caption{(a)~PrOs$_4$As$_{12}$ de Haas-van Alphen frequencices
(paramagnetic metal phase)
as a function of field orientation for rotation
of the sample about the $[1\bar{1}0]$ axis (positive $\theta$ values)
and $[010]$ axis (negative $\theta$ values).
Angles for which the magnetic field lies along crystal axes
are indicated by vertical lines.
Smaller points
show the theoretical frequencies for the
47th and 48th hole-band Fermi surfaces (solid 
and hollow points respectively- see Fig.~\ref{fermifig}).
(b)~LaOs$_4$As$_{12}$ de Haas-van Alphen frequencices
as a function of field orientation for rotation
of the sample about the $[1\bar{1}0]$ axis.
In both (a) and (b), the absence of a symbol
at a particular angle indicates that
that frequency could not be observed or
was indistinct. The inset key indicates
the labelling of the Fermi-surface sections.}
\label{fig3}
\end{figure}

The frequencies
are shown for PrOs$_4$As$_{12}$ (paramagnetic metal phase)
as a function of field orientation in
Fig.~\ref{fig3}(a) and for
LaOs$_4$As$_{12}$ in Fig.~\ref{fig3}(b).
The orientation dependences of the
observable frequencies
suggest that the Fermi surfaces of both materials are rather similar.
In addition, the de Haas-van Alphen frequencies predicted
for the calculated Fermi surface of Fig.~\ref{fermifig}
are shown in Fig.~\ref{fig3}(a).
The LDA/FLAPW calculations also allow one to
deduce effective masses for a particular orbit.
As might be expected,
it is found that the experimentally-observed 
de Haas-van Alphen frequencies
shown in Fig.~\ref{fig3}(a) in general
correspond to the calculated orbits
with the lowest effective masses.

\section{Discussion}
Given the valences of La and Pr,
the similarity of the Fermi surfaces of
PrOs$_4$As$_{12}$ and LaOs$_4$As$_{12}$
suggest that the $4f$-electrons of
the Pr ions make little contribution to the itinerant
quasiparticles in the high-field paramagnetic metal phase
of PrOs$_4$As$_{12}$. This view is also
supported by the strong qualitative similarity between
the theoretical de Haas-van Alphen
frequencies and the experimental data (Fig.~\ref{fig3}(a));
the LDA/FLAPW calculations assume localized $4f$ electrons.

The situation in PrOs$_4$As$_{12}$ may be rather analogous to that
observed in a number of other f-electron systems
(see Ref.~\cite{drymiotis} and citations therein).
It is likely that both the antiferromagnetic 
and paramagnetic metal phases possess
highly-correlated $4f$ electrons,
but with somewhat different effective Kondo
temperatures. In such a scheme, the effective Kondo temperature
of the paramagnetic metal phase will be relatively small
(see the values $T_{\rm K}\sim 1-3.5$~K discussed 
above~\cite{maplepnas}),
so that the properties of the $4f$ electrons will
be almost indistinguishable from those of
localized ionic moments.
By comparison, it is likely that the antiferromagnetic metal
phase will have
a relatively large effective Kondo 
temperature by comparison,
causing the $4f$ electrons to be in the
mixed-valence regime with significant $spd$
and $f$ hybridization at low temperatures.
Consequently one might expect
the charge degrees of freedom
of PrOs$_4$As$_{12}$ in the antiferromagnetic phase
to be describable in terms of
itinerant quasiparticles with a
large effective mass, a view that is
supported by the substantial electronic contribution to the 
heat capacity~\cite{maplepnas}.
Itinerant quasiparticles are 
preferable from a zero-point energy standpoint
at low temperatures,
but the quasi-localised
$4f$ electrons of the paramagnetic metal phase will be
favoured on entropic grounds at elevated 
temperatures and fields~\cite{drymiotis}.
Although the exact details may differ, a
similar picture was recently found to obtain in
CeIn$_3$~\cite{harrisonnature}.
Here, Fermi-surface pockets of very heavy quasiparticles
are observed in the antiferromagnetic phase
at low magnetic fields~\cite{harrisonnature}.
The presence of these itinerant quasiparticles
is able to account for the large value of 
the electronic heat capacity coefficient $\gamma$
in the antiferromagnetic phase.
However, as the field is increased, 
the heavy-quasiparticle pockets are destroyed,
shortly before the field-induced suppression of 
antiferromagnetism.

\section{Summary}
The susceptibility and
de Haas-van Alphen effect have
been measured in single crystals of the 
filled skutterudites PrOs$_4$As$_{12}$ and
LaOs$_4$Sb$_{12}$ using both pulsed and
quasistatic magnetic fields. A cascade of three or four
strongly field-orientation-dependent metamagnetic transitions
is observed in PrOs$_4$As$_{12}$
on sweeping the field from the antiferromagnetic
phase to the paramagnetic metal phase. 
The Fermi-surface topologies of
LaOs$_4$As$_{12}$ and the paramagnetic metal
phase of PrOs$_4$As$_{12}$ 
are found to be very similar. 
In addition, they are
in reasonable agreement with the predictions 
of bandstructure calculations for LaOs$_4$As$_{12}$
on a PrOs$_4$As$_{12}$ lattice.
Both facts suggest that the $4f$ electrons of Pr
are essentially localized in the paramagnetic-metal phase
of PrOs$_4$As$_{12}$. 
However, whilst the properties of LaOs$_4$As$_{12}$
suggest a conventional nonmagnetic Fermi liquid,
the effects of direct exchange and electron
correlations may be detected in the paramagnetic metal
phase of PrOs$_4$As$_{12}$;
a direct exchange energy $\approx 1$~meV
splits the bands, leading to beats in the
de Haas-van Alphen oscillations, and the quasiparticle effective masses
in PrOs$_4$As$_{12}$ are found to decrease with 
increasing field, probably reflecting the gradual suppression of
magnetic fluctuations associated with proximity to the low-temperature,
low-field antiferromagnetic state. 

\section*{Acknowledgements}
Research at UCSD was supported by the U. S. Department
of Energy (DoE) under Grant No. DE-FG02-04ER46105, the
U.S. National Science Foundation (NSF) under Grant No. DMR
0335173, and the National Nuclear Security Administration
under the Stewardship Science Academic Alliances Program
through DOE Research Grant No. DE-FG52-03NA00068.
The work carried out at NHMFL was supported
by DoE (Grant LDRD-DR 20070013)
and by NSF and 
the State of Florida. 
Studies at Kobe are supported by
a Grant-in-Aid for Scientific Research Priority Area
``Skutterudite'' (15072204), MEXT, Japan. 
We thank Neil Harrison, Ross McDonald and Stan
Tozer for experimental assistance and useful discussions.

\end{document}